\documentclass[12pt]{article}
\usepackage{amsfonts}
\usepackage[centertags]{amsmath}
\usepackage{amssymb}
\usepackage[verbose,colorlinks=true,naturalnames=true,linkcolor=blue,]{hyperref}
\newcounter{rown}

\begin{document}

\title{Once more on parastatistics
}
\author{{\it V.N. Tolstoy}\\
\\
Lomonosov Moscow State University, \\
Skobeltsyn Institute of Nuclear Physics,\\
Moscow 119991, Russian Federation}
\date{}
\maketitle

\begin{abstract} Equivalence between algebraic structures generated by parastatistics
triple relations of Green (1953) and  Greenberg -- Messiah (1965), and certain
orthosymplectic $\mathbb{Z}_2\times \mathbb{Z}_2$-graded Lie superalgebras is found
explicitly. Moreover, it is shown that such superalgebras give more complex para-Fermi
and para-Bose systems then ones of Green -- Greenberg -- Messiah.
\end{abstract}

\setcounter{equation}{0}
\section{Introduction}
The usual creation and annihilation operators of identical particles, fermions
$a^{\pm}_i$ ($i=1,\ldots,m$) and bosons $b^{\pm}_j$ ($j=1,\ldots,n$), satisfy the
canonical commutation relations:
\begin{eqnarray}\label{pr1} %
\{a^{\zeta}_{i},a^{\eta}_{j}\}\!\!&=\!\!&\frac{1}{2}|\eta-\zeta|\delta_{ij},\qquad
[b^{\zeta}_{i},b^{\eta}_j]\;\;=\;\;\frac{1}{2}(\eta-\zeta)\delta_{ij}.
\end{eqnarray}%
Here and elsewhere the Greek letters $\zeta,\eta\in\{+,-\}$ if they are upper indexes,
and they are interpreted as +1 and -1 in the algebraic expressions of the type
$\eta-\zeta$.

From the relations (\ref{pr1}) it follows the so-called "{\it symmetrization postulate}"
(SP): {\it States of more than one identical particle must be antisymmetric (fermions) or
symmetric (bosons) under permutations}.

In 1953 Green \cite{Green1953} proposed to refuse SP and he introduced algebras with the
triple relations:
\begin{eqnarray}\label{pr3} %
[[a^{\zeta}_i,a^{\eta}_j],a^{\xi}_{k}]\!\!&=\!\!&|\xi-\eta|\delta_{jk}a^{\zeta}_i-
|\xi-\zeta|\delta_{ik}a^{\eta}_j\quad({\rm parafermions}),
\\[4pt]\label{pr4}
[\{b^{\zeta}_i,b^{\eta}_j\},b^{\xi}_{k}]\!\!&=\!\!&(\xi-\eta)\delta_{jk}b^{\zeta}_i+
(\xi-\zeta)\delta_{ik}b^{\eta}_j\quad({\rm parabosons}).
\end{eqnarray}%
The usual fermions and bosons satisfy these relations but also another solutions exist.

In 1962 Kamefuchi and Takahashi \cite{KamefuchiTakahashi1962} (also see
\cite{RyanSudarshan1963}) shown that the parafermionic algebra is isomorphic to the
orthogonal Lie algebra $\mathfrak{o}(2m+1):=\mathfrak{o}(2m+1,\mathbb{C})$. Later in 1980
Ganchev and Palev \cite{GanchevPalev1980} proved that the parabosonic algebra is
isomorphic to the orthosymplectic $\mathbb{Z}_2$-graded Lie superalgebra
$\mathfrak{osp}(1|2n)$.

In 1965 Greenberg and Messiah \cite{GreenbergMessiah1965} considered parasystem
consisting simultaneously from parafermions and parabosons and they defined the relative
commutation rules between parafermions and parabosons. There are two types of such
relations:
\begin{eqnarray}%
\begin{array}{rclcl}\label{pr7}
[[a_i^{\zeta},a_j^{\eta}],b_k^{\xi}]\!\!&=\!\!&0,\qquad\qquad\qquad\quad
[\{b_i^{\zeta},b_j^{\eta}\},a_k^{\xi}]\!\!&=\!\!&0,
\\[4pt] 
[[a_i^{\zeta},b_j^{\eta}],a_k^{\xi}]\!\!&=\!\!&-|\xi-\zeta|\delta_{ik} b_j^{\eta},
\qquad\{[a_i^{\zeta},b_j^{\eta}],b_k^{\xi}\}\!\!&=\!\!&(\xi-\eta)\delta_{jk}a_i^{\zeta},
\end{array}
\\[7pt]
\begin{array}{rclcl}\label{pr8}
[[a_i^{\zeta},a_j^{\eta}],b_k^{\xi}]\!\!&=\!\!&0,\qquad\qquad\qquad\quad
[\{b_i^{\zeta},b_j^{\eta}\},a_k^{\xi}]\!\!&=\!\!&0,
\\[4pt] 
\{\{a_i^{\zeta},b_j^{\eta}\},a_k^{\xi}\}\!\!&=\!\!&|\xi-\zeta|\delta_{ik} b_j^{\eta},
\qquad\;\;\,
[\{a_i^{\zeta},b_j^{\eta}\},b_k^{\xi}]\!\!&=\!\!&(\xi-\eta)\delta_{jk}a_i^{\zeta},
\end{array}
\end{eqnarray}%
where $i,j,k=1,2,\ldots,m$ for the symbols $a$'s and $i,j,k=1,2,\ldots,n$ for the symbols
$b$'s. The first case (\ref{pr7}) was called as {\it the relative para-Fermi set} and the
second case (\ref{pr8}) was called as {\it the relative para-Boson set}\footnote{The
names {\it the relative para-Fermi and para-Boson set} are directly related to type of
the Lie bracket (commutator or anticommutator) given between parafermion and paraboson
elements.}.

In 1982 Palev \cite{Palev1982} shown that the case (\ref{pr7}) with (\ref{pr3}) and
(\ref{pr4}) is isomorphic to the orthosymplectic $\mathbb{Z}_2$-graded Lie superalgebra
$\mathfrak{osp}(2m+1|2n)$. No any similar solution for the second case (\ref{pr8}) was
known up to now.

Here we show that the case (\ref{pr8}) is isomorphic to the orthosymplectic
$\mathbb{Z}_2\times\mathbb{Z}_2$-graded superalgebra $\mathfrak{osp}(1,2m|2n,0)$.
Moreover it will demonstrate that the more general mixed parasystem, which simultaneously
involves the relative para-Fermi and relative para-Bose sets, contains two sorts of
parafermions and one sort of parabosons and it is isomorphic to the orthosymplectic
$\mathbb{Z}_2\times\mathbb{Z}_2$-graded superalgebra $\mathfrak{osp}(2m_1+1,2m_2|2n,0)$.
All previous cases are particular (degenerated) variants of this general case.

The paper is organized as follows. Section 2 provides a definition and general structure
of $\mathbb{Z}_2\times\mathbb{Z}_2$-graded superalgebras and also a matrix realization
and a Cartan-Weyl basis of the general linear $\mathbb{Z}_2\times\mathbb{Z}_2$-graded
superalgebra $\mathfrak{gl}(m_1^{},m_2^{}|n_1^{},n_2^{})$. In Section 3 we describe the
orthosymplectic $\mathbb{Z}_2\times\mathbb{Z}_2$-graded superalgebra
$\mathfrak{osp}(2m_1+1,2m_2|2n,0)$ and  show that a part of its defining triple relations
in the terms of short root vectors coincides with the the relative para-Bose set.

\setcounter{equation}{0}

\section{Superalgebra $\mathfrak{gl}(m_1^{},m_2^{}|n_1^{},n_2^{})$}

At first we remind a general definition of the $\mathbb{Z}_2\times\mathbb{Z}_2$-graded
superalgebra \cite{RittebergWyler1978a}, \cite{RittenbergWyler1978b}.

The $\mathbb{Z}_2\times\mathbb{Z}_2$-graded LSA $\mathfrak{g}$, as a linear
space, is a direct sum of four graded components
\begin{eqnarray}\label{sasa5}
\mathfrak{g}\!\!&=\!\!&\bigoplus_{\mathbf{a}=(a_1,a_2)}
\mathfrak{g}_{\mathbf{a}}\;= \;\mathfrak{g}_{(0,0)}\oplus
\mathfrak{g}_{(1,1)}\oplus\mathfrak{g}_{(1,0)}\oplus
\mathfrak{g}_{(0,1)}~
\end{eqnarray}
with a bilinear operation  $[\![\cdot,\cdot]\!]$ satisfying the identities (grading,
symmetry, Jacobi):
\begin{eqnarray}\label{sasa6}
\deg([\![x_{\mathbf{a}},y_{\mathbf{b}}]\!])\!\!&=\!\!&\deg(x_{\mathbf{a}})+
\deg(x_{\mathbf{b}})\,=\,\mathbf{a}+\mathbf{b}\,=\,(a_1+b_1,a_2+b_2), 
\\[4pt]\label{sasa7}
[\![x_{\mathbf{a}},y_{\mathbf{b}}]\!]\!\!&=\!\!&-(-1)^{\mathbf{a}\mathbf{b}}
[\![y_{\mathbf{b}},x_{\mathbf{a}}]\!], 
\\[4pt]\label{sasa8}
[\![x_{\mathbf{a}},[\![y_{\mathbf{b}},z]\!]]\!]\!\!&=\!\!&[\![[\![x_{\mathbf{a}},
y_{\mathbf{b}}]\!],z]\!]+ (-1)^{\mathbf{a}\mathbf{b}}
[\![y_{\mathbf{b}},[\![x_{\mathbf{a}},z]\!]]\!],
\end{eqnarray}
where the vector $(a_1+b_1,a_2+b_2)$ is defined $\!\!\!\mod\!(2,2)$ and $\mathbf{ab}=
a_1b_1+a_2b_2$. Here in (\ref{sasa6})-(\ref{sasa8}) $x_{\mathbf{a}}\in
\mathfrak{g}_{\mathbf{a}}$, $x_{\mathbf{b}}\in\mathfrak{g}_{\mathbf{b}}$, and the element
$z\in\mathfrak{g}$ is not necessarily homogeneous. From (\ref{sasa6}) it is
follows that $\mathfrak{g}_{(0,0)}$ is a Lie subalgebra in $\mathfrak{g}$, and
the subspaces $\mathfrak{g}_{(1,1)}$, $\mathfrak{g}_{(1,0)}$ and $\mathfrak{g}_{(0,1)}$
are $\mathfrak{g}_{(0,0)}$-modules. It should be noted that
$\mathfrak{g}_{(0,0)}\oplus\mathfrak{g}_{(1,1)}$ is a Lie subalgebra in
$\mathfrak{g}$ and the subspace $\mathfrak{g}_{(1,0)}\oplus\mathfrak{g}_{(0,1)}$
is a $\mathfrak{g}_{(0,0)}\oplus\mathfrak{g}_{(1,1)}$-module, and moreover
$\{\mathfrak{g}_{(1,1)},\,\mathfrak{g}_{(1,0)}\}\subset\mathfrak{g}_{(0,1)}$ and vice
versa $\{\mathfrak{g}_{(1,1)},\,\mathfrak{g}_{(0,1)}\}\subset\mathfrak{g}_{(1,0)}$. From
(\ref{sasa6}) and (\ref{sasa7}) it is follows that the general Lie bracket
$[\![\cdot,\cdot]\!]$ for homogeneous elements posses two value: commutator
$[\cdot,\cdot]$ and anticommutator $\{\cdot,\cdot\}$ as well as in a case of usual
$\mathbb{Z}_2$-graded Lie superalgebras \cite{Kac1977}.

Now we construct a $\mathbb{Z}_2\times\mathbb{Z}_2$-graded matrix superalgebras
$\mathfrak{gl}(m_1^{},m_2^{}|n_1^{},n_2^{})$.

Let an arbitrary
$(m_1^{}+m_2^{}+n_1^{}+n_2^{})\times(m_1^{}+m_2^{}+n_1^{}+n_2^{})$-matrix $M$ be
presented in the following block form\footnote{It is evidently supposed that all such
matrices in each block-row or in each block-column have the same number of rows or
columns}:
\begin{eqnarray}\label{gl9}
M\!\!&=\!\!&\left(\!\!%
\begin{array}{cccc}
  A_{(0,0)} & A_{(1,1)} & A_{(1,0)} & A_{(0,1)} \\
  B_{(1,1)} & B_{(0,0)} & B_{(0,1)} & B_{(1,0)} \\
  C_{(1,0)} & C_{(0,1)} & C_{(0,0)} & C_{(1,1)} \\
  D_{(0,1)} & D_{(1,0)} & D_{(1,1)} & D_{(0,0)} \\
\end{array}%
\!\!\!\right),
\end{eqnarray} 
where the diagonal block matrices $A_{(0,0)},B_{(0,0)},C_{(0,0)},D_{(0,0)}$ have the
dimensions $m_1^{}\times m_1^{}$, $m_2^{}\times m_2^{}$, $n_1^{}\times n_1^{}$ and
$n_2^{}\times n_2^{}$ correspondingly, the dimensions of the non-diagonal block matrices
$A_{(1,1)},A_{(1,0)},A_{(0,1)}$, etc. are easy determined by the dimensions of these
diagonal block matrices. The matrix $M$ can be split into the sum of four matrices:
\begin{eqnarray}\label{gl10}
M\!\!&=\!\!&M_{(0,0)}^{}+M_{(1,1)}^{}+M_{(1,0)}^{}+M_{(0,1)}^{}\;=
\\[3pt]\nonumber
\!\!&=\!\!&\left(\!\!
\begin{array}{cccc}
  A_{(0,0)} & 0 & 0 & 0 \\
  0 & B_{(0,0)} & 0 & 0 \\
  0 & 0 & C_{(0,0)} & 0 \\
  0 & 0 & 0 & D_{(0,0)} \\
\end{array}%
\!\!\!\right)+\left(\!\! %
\begin{array}{cccc}
  0 & A_{(1,1)} & 0 & 0 \\
  B_{(1,1)} & 0 & 0 & 0 \\
  0 & 0 & 0 & C_{(1,1)} \\
  0 & 0 & D_{(1,1)} & 0 \\
\end{array}%
\!\!\!\right)
\\[3pt]\nonumber
\!\!&+\!\!&\left(\!\!
\begin{array}{cccc}
  0 & 0 & A_{(1,0)} & 0 \\
  0 & 0 & 0 & B_{(1,0)} \\
  C_{(1,0)} & 0 & 0 & 0 \\
  0 & D_{(1,0)} & 0 & 0 \\
\end{array}%
\!\!\!\right)+\left(\!\! %
\begin{array}{cccc}
  0 & 0 & 0 & A_{(0,1)} \\
  0 & 0 & B_{(0,1)} & 0 \\
  0 & C_{(0,1)} & 0 & 0 \\
  D_{(0,1)} & 0 & 0 & 0 \\
\end{array}%
\!\!\!\right).
\end{eqnarray}
Let us define the general commutator $[\![\cdot,\cdot]\!]$ on a space of all such
matrices by the following way:
\begin{eqnarray}\label{gl11}
[\![M_{(a_1^{},a_2^{})},M_{(b_1^{},b_2^{})}']\!]\!\!&:=\!\!&M_{(a_1^{},a_2^{})}
M_{(b_1^{},b_2^{})}'-(-1)^{a_1^{}b_1^{}+a_2^{}b_2^{}}M_{(b_1^{},b_2^{})}
'M_{(a_1^{},a_2^{})}.
\end{eqnarray}
for the homogeneous components  $M_{(a_1,a_2)}$ and $M_{(b_,b_2)}$. For arbitrary
matrices $M$ and $M'$ the commutator $[\![\cdot,\cdot]\!]$ is extended by linearity. It
is easy to check that
\begin{eqnarray}\label{gl12}
[\![M_{(a_1,a_2)},M_{(b_1,b_2)}']\!]\!\!&=\!\!&M_{(a_1+a_2,b_1+b_2)}'',
\end{eqnarray}
where the sum $(a_1+a_2,b_1+b_2)$ is defined $\!\!\!\!\mod\!(2,2)$. Thus the {\it
grading} condition (\ref{sasa6}) is available. The {\it symmetry} and {\it Jacobi}
identities (\ref{sasa7}) and (\ref{sasa8}) are available too. Hence we obtain a Lie
superalgebra which is called $\mathfrak{gl}(m_1^{},m_2^{}|n_1^{},n_2^{})$. It should be
noted that
\begin{eqnarray}\label{gl13}
\begin{array}{rcl}
[\![M_{\mathbf{a}}^{},M_{\mathbf{b}}']\!]\!\!&=\!\!&[M_{\mathbf{a}},M_{\mathbf{b}}']\quad\;\;
{\rm if}\;\;\;\mathbf{a}\mathbf{b}=0,2,
\\[7pt]
[\![M_{\mathbf{a}}^{},M_{\mathbf{b}}']\!]\!\!&=\!\!&\{M_{\mathbf{a}}^{},
M_{\mathbf{b}}'\}\quad {\rm if}\;\;\mathbf{a}\mathbf{b}=1.
\end{array}
\end{eqnarray}%

Now we consider the Cartan-Weyl basis of $\mathfrak{gl}(m_1^{},m_2^{}|n_1^{},n_2^{})$ and
its supercommutation ($\mathbb{Z}_2\times\mathbb{Z}_2$-graded) relations. In accordance
with the block structure of the $\mathbb{Z}_2\times\mathbb{Z}_2$-graded matrix
(\ref{gl9}) we introduce a $\mathbb{Z}_2\times\mathbb{Z}_2$-graded function (grading)
$\rm{\mathbf{d}}(\cdot)$ defined on the integer segment
$[1,2,\ldots,m_1^{},m_1^{}+1,\ldots,m_1^{}+m_2^{},m_1^{}+m_2^{}+1,\ldots,m_1^{}+m_2^{}+n_1^{},
m_1^{}+m_2^{}+n_1^{}+1,\ldots,m_1^{}+m_2^{}+n_1^{}+n_2^{}]$ as follows:
\begin{eqnarray}\label{gl14}
{\rm\mathbf{d}}_{i}\!\!&:=\!\!&{\rm\mathbf{d}}(i)\;\;=\;\;\left\{ %
\begin{array}{rcl}(0,0)\;\;\;{\rm{for}}\;\;i\!\!&=\!\!&1,2,\ldots,m_1^{},\\[2pt]
(1,1)\;\;\;{\rm{for}}\;\;i\!\!&=\!\!&m_1^{}+1,\ldots,m_1^{}+m_2^{},\\[2pt]
(1,0)\;\;\;{\rm{for}}\;\;i\!\!&=\!\!&m_1^{}+m_2^{}+1,\ldots,m_1^{}+m_2^{}+n_1^{},\\[2pt]
(0,1)\;\;\;{\rm{for}}\;\;i\!\!&=\!\!&m_1^{}+m_2^{}+n_1^{}+1,\ldots,m_1^{}+m_2^{}+n_1^{}+
n_2^{}.\\
\end{array}\right.
\end{eqnarray}
Let $e_{ij}$ be the $(m_1^{}+m_2^{}+n_1^{}+n_2^{})\times(m_1^{}+m_2^{}+n_1^{}+n_2^{})$
matrix (\ref{gl9}) with 1 is in the $(i,j)$-th place and other entries 0. The matrices
$e_{ij}$ $(i,j=1,2,\ldots,m_1^{}+m_2^{}+n_1^{}+n_2^{})$ are homogeneous, moreover, the
grading $\deg(e_{ij})$ is determined by
\begin{eqnarray}\label{gl15} %
\deg(e_{ij})\!\!&=\!\!&{\rm\mathbf{d}}_{ij}\;:=\;{\rm\mathbf{d}}_{i}+
{\rm\mathbf{d}}_{j}\quad(\!\!\!\!\!\mod\!(2,2)\,),
\end{eqnarray}
and the supercommutator for such matrices is given as follows
\begin{eqnarray}\label{gl16} %
[\![e_{ij},\,e_{kl}]\!]\!\!&:=\!\!&e_{ij}e_{kl}-(-1)^{{\rm\mathbf{d}}_{ij}
{\rm\mathbf{d}}_{kl}}e_{kl}e_{ij}.
\end{eqnarray}
It is easy to check that
\begin{eqnarray}\label{gl17} %
[\![e_{ij},\,e_{kl}]\!]\!\!&=\!\!&\delta_{jk}e_{il}-
(-1)^{{\rm\mathbf{d}}_{ij}{\rm\mathbf{d}}_{kl}}\delta_{il}e_{kj}.
\end{eqnarray}
The elements $e_{ij}$ $(i,j=1,2,\ldots,m_1^{}+m_2^{}+n_1^{}+n_2^{})$ with the relations
(\ref{gl17}) generates the Lie superalgebra $\mathfrak{gl}(m_1^{},m_2^{}|n_1^{},n_2^{})$.
The elements $h_{i}:=e_{ii}$ $(i,j=1,2,\ldots,m_1^{}+m_2^{}+n_1^{}+n_2^{})$ compose a
basis in the Cartan subalgebra $\mathfrak{h}(m_1^{}+m_2^{}|n_1^{}+n_2^{})\subset
\mathfrak{gl}(m_1^{},m_2^{}|n_1^{},n_2^{})$.

The Lie superalgebra $\mathfrak{gl}(m_1^{},m_2^{}| n_1^{},n_2^{}$ play a special role
among all finite dimensional $\mathbb{Z}_2\times \mathbb{Z}_2$-graded Lie superalgebras.
Namely, a general Ado's theorem is valid. It states: {\it Any finite dimensional Lie
$\mathbb{Z}_2\times \mathbb{Z}_2$-graded superalgebra can be realized in terms of a
subalgebra of $\mathfrak{gl}(m_1^{},m_2^{}|n_1^{},n_2^{})$}. This theorem was proved by
Scheunert \cite{Scheunert1979} for all finite dimensional graded generalized Lie algebras
including our cases.

As an illustration of the Ado's theorem, in the next section we give realization of the
orthosymplectic $\mathbb{Z}_2\times \mathbb{Z}_2$-graded superalgebra
$\mathfrak{osp}(2m_1^{}+1,2m_2^{}|2n,0)$ in terms of the superalgebra
$\mathfrak{gl}(2m_1^{}+1,2m_2^{}|2n,0)$ and, moreover, we present a Cartan-Weyl basis of
the orthosymplectic superalgebra and its explicit commutation relations and we also show
that a subset of the short root vectors of the Cartan-Weyl basis generates this
superalgebra and describe the parastatistics with the relative para-Fermi and para-Bose
sets simultaneously.

\setcounter{equation}{0}
\section{Orthosymplectic superalgebra $\mathfrak{osp}(2m_1+1,2m_2|2n,0)$ and
its relation with parastatistics 
}%
We start with an explicit description of embedding of the orthosymplectic Lie
superalgebra $\mathfrak{osp}(2m_1+1,2m_2|2n,0)$ in the general linear Lie superalgebra
$\mathfrak{gl}(2m_1+1,2m_2|2n,0)$. For this propose the $\mathbb{Z}_2\times
\mathbb{Z}_2$-graded integer segment $\mathbb{S}_N^{(\rm{\mathbf{d}})}\,:=\,[1,2,\ldots,
2N+1]$, where $N=m_1+m_2+n$, with the grading $\rm{d}(\cdot)$ given by
\begin{eqnarray}\label{orn1}
{\rm\mathbf{d}}_{i}\!\!&:=\!\!&{\rm\mathbf{d}}(i)\;\;=\;\;\left\{ %
\begin{array}{rcl}(0,0)\;\;\;{\rm{for}}\;\;i\!\!&=\!\!&1,2,\ldots,2m_1^{},\\[2pt]
(1,1)\;\;\;{\rm{for}}\;\;i\!\!&=\!\!&2m_1^{}+1,\ldots,2m_1^{}+2m_2^{},\\[2pt]
(1,0)\;\;\;{\rm{for}}\;\;i\!\!&=\!\!&2m_1^{}+2m_2^{}+1,\ldots,2m_1^{}+
2m_2^{}+2n. 
\end{array}\right.
\end{eqnarray}
is reindexed by the following way $\tilde{\mathbb{S}}_N^{(\rm{\mathbf{d}})}\,:=
\,[0,\pm1,\pm2,\ldots,\pm{N}]$ with the grading $\rm{\mathbf{d}}(\cdot)$ given by
\begin{eqnarray}\label{orn2}
{\rm\mathbf{d}}_{i}\!\!&:=\!\!&{\rm\mathbf{d}}(i)\;\;=\;\;\left\{\begin{array}{rcl}
(0,0)\;\;\;{\rm{for}}\;\;i\!\!&=\!\!&0,\pm1,\pm2,\ldots,\pm m_1^{},\\[2pt]
(1,1)\;\;\;{\rm{for}}\;\;i\!\!&=\!\!&\pm(m_1^{}+1),\ldots,\pm(m_1^{}+m_2^{}),\\[2pt]
(1,0)\;\;\;{\rm{for}}\;\;i\!\!&=\!\!&\pm(m_1^{}+m_2^{}+1).
\ldots,\pm(m_1^{}+m_2^{}+n), 
\end{array}\right.
\end{eqnarray}
Rows and columns of $\mathbb{Z}_2\times\mathbb{Z}_2$-graded $(2N+1)\times(2N+1)$-matrices
are enumerated by the indices $0,1,-1,2,-2,\ldots,\ldots,N,-N$ ($N=m_1+m_2+n$). Let
$e_{ij}$($i,j\in\tilde{\mathbb{S}}_N^{(\rm{\mathbf{d}})}$) be the standard (unit) basis
of $\mathfrak{gl}(2m_1+1,2m_2|2n,0)$ with the given indexing and the canonical
supercommutation relations:
\begin{eqnarray}\label{orn3} %
[\![e_{ij},\,e_{kl}]\!]\!\!&=\!\!&\delta_{jk}e_{il}-(-1)^{{\rm{\mathbf{d}}}_{ij}
{\rm{\mathbf{d}}}_{kl}} \delta_{il}e_{kj},
\end{eqnarray}
where ${\rm{\mathbf{d}}}_{ij}={\rm{\mathbf{d}}}_{i}+{\rm{\mathbf{d}}}_{j}$ and the
grading ${\rm{\mathbf{d}}}(\cdot)$ is given by (\ref{orn2}).

The orthosymplectic ($\mathbb{Z}_2\times\mathbb{Z}_2$-graded) Lie superalgebra
$\mathfrak{osp}(2m_1+1,2m_2|2n,0)$ is embedded in $\mathfrak{gl}(2m_1+1,2m_2|2n,0)$ as a
linear span of the elements
\begin{eqnarray}\label{orn4} %
x_{ij}\!\!&:=\!\!&e_{i,-j}-(-1)^{{\rm{\mathbf{d}}}_{i}{\rm{\mathbf{d}}}_{j}+
{\rm{\mathbf{d}}}_{ij}^2}\phi_{i}\phi_{j}e_{j,-i}\qquad
(i,j\in\tilde{\mathbb{S}}_N^{(\rm{\mathbf{d}})}),
\end{eqnarray}
where the index function $\phi_i$ is given as follows
\begin{eqnarray}\label{orn5}
\phi_{i}\!\!&:=\!\!\left\{ %
\begin{array}{rcl}1\;\;\;{\rm{if}}\;\;i\!\!&=\!\!&0,\pm1,\pm2,\ldots,\pm(m_1+m_2),\\[2pt]
1\;\;\;{\rm{if}}\;\;i\!\!&=\!\!&m_1+m_2+1,\ldots,m_1+m_2+n,\\[2pt]
-1\;\;\;{\rm{if}}\;\;i\!\!&=\!\!&-m_1-m_2-1,\ldots,-m_1-m_2-n.
\end{array}\right.
\end{eqnarray}
It is easy to verify that the the elements (\ref{orn4}) satisfy the following
supercommutation relations
\begin{eqnarray}\label{orn6} %
\begin{array}{rcl}
[\![x_{ij},\,x_{kl}]\!]\!\!&=\!\!&\delta_{j,-k}x_{il}-
\delta_{j,-l}(-1)^{{\rm{\mathbf{d}}}_{k}{\rm{\mathbf{d}}}_{l}+{\rm\mathbf{d}}_{kl}^2}
\phi_k\phi_lx_{ik}
\\[9pt] 
&&-\delta_{i,-k}(-1)^{{\rm{\mathbf{d}}}_{i}{\rm{\mathbf{d}}}_{j}+{\rm\mathbf{d}}_{ij}^2}
\phi_i\phi_jx_{jl}-\delta_{i,-l}(-1)^{{\rm\mathbf{d}}_{ij}{\rm{\mathbf{d}}}_{ik}}x_{kj}.
\end{array}
\end{eqnarray}
Not all elements (\ref{orn4}) are linearly independent because they satisfy the relations
\begin{eqnarray}\label{orn7} %
x_{ij}\!\!&=\!\!&-(-1)^{{\rm{\mathbf{d}}}_{i}{\rm{\mathbf{d}}}_{j}+{\rm\mathbf{d}}_{ij}^2}
\phi_{i}\phi_{j}x_{ji}\qquad (i,j\in\tilde{\mathbb{S}}_N^{(\rm{\mathbf{d}})}),
\end{eqnarray}
and what is more
\begin{eqnarray}\label{orn8} %
x_{ii}\!\!&=\!\!&0\qquad {\rm for}\;\;i=0,\pm1,\pm2,\ldots,\pm(m_1+m_2).
\end{eqnarray} 
From the general supercommutation relations (\ref{orn6}) it follows at once that the
short root vectors $x_{0i}$ ($i=\pm1,\pm2,\dots,\pm(m_1+m_2+n)$) satisfy the following
triple relations:
\begin{eqnarray}\label{orn17} 
[\![[\![x_{0i},\,x_{0j}]\!],x_{0k}]\!]\!\!&=\!\!&-\delta_{j,-k}\phi_{j}x_{0i}+
\delta_{i,-k}(-1)^{{\rm{\mathbf{d}}}_{i}{\rm{\mathbf{d}}}_{j}}\phi_{i}x_{0j}.
\end{eqnarray}
Conversely, let the abstract $\mathbb{Z}_2\times\mathbb{Z}_2$-graded generators $x_{0i}$
($i=\pm1,\pm2,\dots,\pm(m_1+m_2+n)$) with the grading $\deg(x_{0i})=\rm{\mathbf{d}}_{0i}
\equiv\rm{\mathbf{d}}_{0}+\rm{\mathbf{d}}_{i}=\rm{\mathbf{d}}_{i}$, where
$\rm{\mathbf{d}}_{i}$ is given by (\ref{orn2}), satisfy the relations (\ref{orn17}),
where the the index function $\phi_{i}$ is determined by (\ref{orn5}), then it is not
difficult to check that these relations generate for the superalgebra
$\mathfrak{osp}(2m_1+1,2m_2|2n,0)$.

The defining relations (\ref{orn17}) can be rewritten in detailed in accordance with the
explicit grading of their generators using the following notations\footnote{Here anywhere
$\zeta,\eta,\xi\in\{+,-\}$.}:
\begin{eqnarray}\label{orn21} 
\begin{array}{rclcl}
a_{i}^{-\zeta}\!\!&:=\!\!&\zeta\sqrt{2}\,{x}_{0,\zeta{i}}\qquad\qquad
(\deg(a_{i}^{\zeta})\!\!&=\!\!&(0,0))\quad\;{\rm for}\;\;i =1,2,\ldots,m_1,
\\[4pt]
\tilde{a}_{i}^{-\zeta}\!\!&:=\!\!&\zeta\sqrt{2}\,{x}_{0,\zeta(m_1+i)}\qquad
(\deg(\tilde{a}_{i}^{\zeta})\!\!&=\!\!&(1,1))\quad\;{\rm for}\;\;i=1,2,\ldots,m_2, 
\\[4pt]
b_{i}^{-\zeta}\!\!&:=\!\!&\zeta\sqrt{2}\,{x}_{0,\zeta(m+i)}\qquad\;\;
(\deg(b_{i}^{\zeta})\!\!&=\!\!&(1,0))\quad\;{\rm for}\;\;i=1,2,\ldots,n, %
\end{array}
\end{eqnarray}
where $m:=m_1+m_2$. Substituting (\ref{orn21}) in (\ref{orn17}) we obtain the different
types of defining triple relations.
\begin{description}
\item[{\rm 1. Parafermion relations}{\rm:}] 
\item[\it{(a1) the defining relations of $\mathfrak{o}(2m_1+1)$} {\rm:}]
\begin{eqnarray}\label{orn25}
[[a_{i}^{\zeta},a_{j}^{\eta}],a_{k}^{\xi}]\!\!&=\!\!&|\xi-\eta|\delta_{jk}a_{i}^{\zeta}-
|\xi-\zeta|\delta_{ik}a_{j}^{\eta}\quad\;{\rm for}\;\; i,j,k\;=\;1,2,\ldots,m_1;
\end{eqnarray}
\item[\it{(a2) the defining  relations of $\mathfrak{o}(2m_2+1)$} {\rm:}]
\begin{eqnarray}\label{orn26}
[[\tilde{a}_{i}^{\zeta},\tilde{a}_{j}^{\eta}],\tilde{a}_{k}^{\xi}]\!\!&=\!\!&|\xi-\eta|
\delta_{jk}\tilde{a}_{i}^{\zeta}-|\xi-\zeta|\delta_{ik}\tilde{a}_{j}^{\eta}\quad\;{\rm
for}\;\; i,j,k\;=\;1,2,\ldots,m_2;
\end{eqnarray}
\item[\it{(a3) the mixed parafermion relations} {\rm:}]
\begin{eqnarray}\label{orn27}
[[a_{i}^{\zeta},a_{j}^{\eta}],\tilde{a}_{k}^{\xi}]\!\!&=\!\!&0,\qquad\qquad\qquad\quad\,
[[\tilde{a}_{i}^{\zeta},\tilde{a}_{j}^{\eta}],a_{k}^{\xi}]\;\;=\;\;0,
\\[7pt]\label{orn28}
[[a_{i}^{\zeta},\tilde{a}_{j}^{\eta}],a_{k}^{\xi}]\!\!&=\!\!&-|\xi-\zeta|
\delta_{ik}\tilde{a}_{j}^{\eta},\qquad
[[a_{i}^{\zeta},\tilde{a}_{j}^{\eta}],\tilde{a}_{k}^{\xi}]\;\;=\;\;|\xi-\eta|
\delta_{jk}a_{i}^{\zeta},
\end{eqnarray}
where $i,j,k=1,\ldots,m_1$ for the symbols $a$'s and $i,j,k=1,\ldots,m_2$ for the symbols
$b$'s.
\item[{\rm 2. Paraboson relations}{\rm:}] %
\item[\it{(b1) the defining  relations of $\mathfrak{osp}(1|2n)$}{\rm:}]
\begin{eqnarray}\label{orn29} 
[\{b_{i}^{\zeta},b_{j}^{\eta}\},b_{k}^{\xi}]\!\!&=\!\!&
(\xi-\eta)\delta_{jk}b_{i}^{\zeta}+ (\xi-\zeta)\delta_{ik}b_{j}^{\eta}
\quad\;{\rm for}\;\; i,j,k\;=\;1,2,\ldots,n.
\end{eqnarray}
\item[{\rm 2. Mixed parafermion and paraboson relations}{\rm:}]
\item[\it{(ab1) the relative para-Fermi set}{\rm:}]
\begin{eqnarray} 
\begin{array}{rclcl}\label{orn33} %
[[a_i^{\zeta},a_j^{\eta}],b_k^{\xi}]\!\!&=\!\!&0,\qquad\qquad\qquad\quad
[\{b_i^{\zeta},b_j^{\eta}\},a_k^{\xi}]\!\!&=\!\!&0,
\\[7pt] 
[[a_i^{\zeta},b_j^{\eta}],a_k^{\xi}]\!\!&=\!\!&-|\xi-\zeta|\delta_{ik} b_j^{\eta},
\qquad\{[a_i^{\zeta},b_j^{\eta}],b_k^{\xi}\}\!\!&=\!\!&(\xi-\eta)\delta_{jk}a_i^{\zeta},
\end{array}
\end{eqnarray}%
where $i,j,k=1,2,\ldots,m_1$ for the symbols $a$'s and $i,j,k=1,\ldots,n$ for the symbols
$b$'s; 
\item[\it{(ab2) the relative para-Bose set}{\rm:}]
\begin{eqnarray}  
\begin{array}{rclcl}\label{orn35} %
[[\tilde{a}_i^{\zeta},\tilde{a}_j^{\eta}],b_k^{\xi}]\!\!&=\!\!&0,\qquad\qquad\qquad\quad
[\{b_i^{\zeta},b_j^{\eta}\},\tilde{a}_k^{\xi}]\!\!&=\!\!&0,
\\[7pt] 
\{\{\tilde{a}_i^{\zeta},b_j^{\eta}\},\tilde{a}_k^{\xi}\}\!\!&=\!\!&|\xi-\zeta|\delta_{ik}
b_j^{\eta},
\qquad[\{\tilde{a}_i^{\zeta},b_j^{\eta}\},b_k^{\xi}]\!\!&=\!\!&(\xi-\eta)\delta_{jk}
\tilde{a}_i^{\zeta}, 
\end{array}
\end{eqnarray}%
where $i,j,k=1,2,\ldots,m_2$ for the symbols $\tilde{a}$'s and $i,j,k=1,\ldots,n$ for the
symbols $b$'s;
\item[\it{(ab3) the relations with distinct grading elements}{\rm:}]
\begin{eqnarray}\label{orn37} 
\{[a_i^{\zeta},\tilde{a}_j^{\eta}],b_k^{\xi}\}\!\!&=\!\!&[\{\tilde{a}_j^{\eta},b_k^{\xi}\},
a_i^{\zeta}]\;\;=\;\;\{[b_k^{\xi},a_i^{\zeta}],\tilde{a}_j^{\eta}\}\;\;=\;\;0.
\end{eqnarray}%
\end{description}
The result connected with the relation (\ref{orn17}) can be reformulated the following
way. {\it If we have two sorts of the parafermions $a_i^{\zeta}$ ($i=1,2,\ldots,m_1$) and
$\tilde{a}_i^{\zeta}$ ($i=1,2,\ldots,m_2$) with the triple relations
(\ref{orn25})--(\ref{orn28}) and one sort of the parabosons $b_i^{\zeta}$
($i=1,2,\ldots,n$) with the triple relations (\ref{orn29}) which together satisfy the
relative para-Fermi set (\ref{orn33}) and relative para-Bose set (\ref{orn35}), and they
obey also the triple relations of the form (\ref{orn37}) then this parasystem generate
the orthosymplectic $\mathbb{Z}_2\times\mathbb{Z}_2$-graded Lie superalgebra
$\mathfrak{osp}(2m_1+1,2m_2|2n,0)$}.

We consider two particular cases which are degenerations of
$\mathfrak{osp}(2m_1+1,2m_2|2n,0)$.
\begin{itemize}
\item
If the parasystem consist of only one sort of the parafermions $a_i^{\zeta}$
($i=1,2,\ldots,m_1$) and one sort of the parabosons $b_i^{\zeta}$ ($i=1,2,\ldots,n$) then
we have the parasystem with the relative Fermi set and it generates the orthosymplectic
$\mathbb{Z}_2$-graded Lie superalgebra $\mathfrak{osp}(2m_1+1|2n)=
\mathfrak{osp}(2m_1+1,0|2n,0)$. 
\item
If the parasystem contains one sort of the parafermions $\tilde{a}_i^{\zeta}$
($i=1,2,\ldots,m_2$) and one sort of the parabosons $b_i^{\zeta}$ ($i=1,2,\ldots,n$)
then we have the case of a parasystem with the relative Bose set (see the relations
(\ref{pr3}), (\ref{pr4}) and (\ref{pr8})) and it generates the orthosymplectic
$\mathbb{Z}_2\times\mathbb{Z}_2$-graded  Lie superalgebra 
$\mathfrak{osp}(1,2m_2|2n,0)$.
\end{itemize}
Thus we shown that the para-Fermi and para-Boose triple relations (\ref{pr3}),
(\ref{pr4}) together with the relative para-Bose set (\ref{pr8}) generate the
orthosymplectic $\mathbb{Z}_2\times\mathbb{Z}_2$-graded Lie superalgebra
$\mathfrak{osp}(1,2m|2n,0)$. Moreover, it was shown that the superalgebras
$\mathfrak{osp}(m_1,2m_2|2n,0)$ give more complex para-Fermi and para-Bose system which
contain the relative para-Fermi and para-Bose sets simultaneously.

It should be noted that, probably, for the first time the $\mathbb{Z}_2\times
\mathbb{Z}_2$-graded structure of the relative para-Bose set (\ref{pr8}) was observable
in \cite{YangJing2001a}, \cite{YangJing2001b} (also see \cite{KanakoglouDasHer2009}).

It should be also noted that the obtained relation between the parastatistics and the
orthosymplectic superalgebras allows to apply all mathematical power of the
representation theory of the superalgebras for a detailed description of the
parastatistics, e.g. their Fock spaces etc. (for example, see
\cite{LievensStoilovaJeugt2008} -- \cite{Stoilova2013}).
\subsection*{Acknowledgments} 
The author would like to thank the Organizers for the kind invitation to speak at the
International Workshop 'Supersymmetries and Quantum Symmetries', Dubna, July 29 - August
3, 2013. The paper was supported by the RFBR grant No.11-01-00980-a and the grant
No.12-09-0064 of the Academic Fund Program of the National Research University Higher
School of Economics.

\end{document}